# Lifespan tree of brain anatomy: diagnostic values for motor and cognitive neurodegenerative diseases

Pierrick Coupé[1] (0000-0003-2709-3350), Boris Mansencal[1], José V. Manjón[2], Patrice Péran[3] (0000-0001-7200-0139), Wassilios G. Meissner[4,5,6] (0000-0003-2172-7527), Thomas Tourdias[7,8] (0000-0002-7151-6325), Vincent Planche[4] (0000-0003-3713-227X)

[1] CNRS, Univ. Bordeaux, Bordeaux INP, LABRI, UMR5800, F-33400 Talence, France
[2] ITACA, Universitat Politècnica de València, 46022 Valencia, Spain
[3] ToNIC, Toulouse NeuroImaging Center, Université de Toulouse, Inserm, UPS, Toulouse, France
[4] CHU Bordeaux, Service de Neurologie des Maladies Neurodégénératives, IMNc, F-33000 Bordeaux, France
[5] Univ. de Bordeaux, CNRS, IMN, UMR 5293, F-33000 Bordeaux, France
[6] Dept. Medicine, University of Otago, Christchurch, New Zealand, and New Zealand Brain Research Institute, Christchurch, New Zealand
[7] Inserm U1215 - Neurocentre Magendie, Bordeaux F-33000, France
[8] Service de neuroimagerie, CHU de Bordeaux, F-33000 Bordeaux.

## Abstract

The differential diagnosis of neurodegenerative diseases, characterized by overlapping symptoms, may be challenging. Brain imaging coupled with artificial intelligence has been previously proposed for diagnostic support, but most of these methods have been trained to discriminate only isolated diseases from controls. Here, we develop a novel machine learning framework, named *lifespan tree* of brain anatomy, dedicated to the differential diagnosis between multiple diseases simultaneously. It integrates the modeling of volume changes for 124 brain structures during the lifespan with non-linear dimensionality reduction and synthetic sampling techniques to create easily interpretable representations of brain anatomy over the course of disease progression. As clinically relevant proof-of-concept applications, we constructed a *cognitive lifespan tree* of brain anatomy for the differential diagnosis of six causes of neurodegenerative dementia and a *motor lifespan tree* of brain anatomy for the differential diagnosis of four causes of parkinsonism using 37594 MRI as a training dataset. This original approach enhanced significantly the efficiency of differential diagnosis in the external validation cohort of 1754 cases, outperforming existing state-of-the art machine learning techniques. *Lifespan tree* holds promise as a valuable tool for differential diagnostic in relevant clinical conditions, especially for diseases still lacking effective biological markers.

# 1 Introduction

In the field of neurodegenerative diseases, differential diagnosis may be challenging due to overlapping symptoms between different disorders. Accurate diagnosis is however essential for prognosis, patient and caregiver information, therapeutic management, and enrolment in clinical trials targeting specific disease mechanisms. Except in the particular case of Alzheimer's disease (AD) where pathophysiological biomarkers are available[1] (highlighting tau and amyloid pathology through biofluid analysis and/or positron emission tomography), there is no fully established biomarkers for the diagnosis of frontotemporal dementia (FTD), progressive supranuclear palsy (PSP), dementia with Lewy bodies (DLB), Parkinson's disease (PD) or multiple system atrophy (MSA). Magnetic Resonance Imaging (MRI) plays a pivotal role in the differential diagnosis of these disorders, providing support in some diagnostic criteria[2–4]. MRI provides detailed representation of the brain, enabling the visualization and measurement of cerebral abnormalities linked to neurodegenerative diseases, such as focal atrophy or suggestive signal abnormalities[5]. However, abnormalities can sometimes be subtle or absent, and different disorders can exhibit similar atrophic patterns, complicating definitive diagnosis.

Advancements in diagnostic technology, including the integration of Artificial Intelligence (AI) with imaging techniques, may improve the accuracy and efficiency of diagnosis[6]. AI algorithms can analyze complex imaging data, identify subtle patterns, and differentiate between conditions with high precision. The synergy of MRI and AI holds significant promise for the early and accurate detection of neurological disorders. AI-enhanced MRI analysis may not only improve diagnostic accuracy but also may reduce the time required for interpretation, facilitating decision-making and hopefully, improving patient outcomes[7,8].

To date, AI-enhanced MRI analysis has primarily been used for binary classification tasks, distinguishing patients with one pathology from control cases[8,9]. This application is not useful for clinicians, who generally have no doubts about distinguishing a patient with a neurodegenerative disease from a healthy subject. However, neurologists can face difficulties in the differential diagnosis between different neurodegenerative diseases that share common symptoms and common biomarkers (such as dopaminergic denervation on dopamine transporter imaging, which can be found in all causes of parkinsonism). In this context, the application of AI-enhanced MRI analysis for differential diagnosis, or multiclass classification that differentiates between several disorders, is promising but remains underexplored. Challenges include the complexity of differentiating multiple diseases simultaneously, the lack of large datasets for training models (especially for recent deep learning methods), and the interpretability of the provided decisions, which complicates its integration into clinical routine.

To address the challenge of automatic differential diagnosis, the vast majority of previous studies have employed machine learning techniques, such as Support Vector Machines (SVM), on meaningful features like brain structure volumes[10–17]. Such a framework mitigates the need for large databases[18] and may provide explainable decisions through *post-hoc* analysis of the different features' importance. However, SVM may be suboptimal as the number of classes increases.

Recent research has proposed lifespan modeling of brain structures as an effective method for examining the progression of cerebral atrophy in neurodegenerative diseases[19–23]. This approach can generate brain charts, or trajectories, depicting the volumes of brain structures across the entire lifespan, including preclinical stages. Such lifespan modeling of manually selected structures has been used for the binary classification of AD, demonstrating performance comparable to recent deep learning techniques[24]. However, such a framework cannot be used when considering multiple diseases simultaneously since structures are differentially vulnerable according to the type of disease.

To address the limitations of current methods, we introduce in this paper a novel framework called *lifespan tree*. This framework integrates lifespan modelling of brain volumes derived from MRI data[19–23] with non-linear dimensionality reduction[25] and synthetic resampling techniques. The proposed approach was designed to improve multi-class classification of high-dimensional temporal data, especially when the data are irregularly sampled over time and exhibit significant class imbalance. Moreover, this innovative strategy produces a compact and elegant representation of human brain anatomy along the disease progression allowing for the simultaneously analysis of multiple disorders. Notably, the proposed *lifespan tree* improves the efficiency and interpretability of differential diagnosis.

Herein, we applied our *lifespan tree* on brain anatomy to the challenging problem of differential diagnosis of cognitive and motor impairments that can be caused by multiple neurodegenerative diseases. We first constructed a *cognitive tree* with seven branches to distinguish normal aging from six diseases or syndromes: AD, DLB, PSP, the behavioral variant of FTD (bvFTD), progressive non-fluent aphasia (PNFA) and semantic dementia (SD). We demonstrated the superiority of our method compared to the state-of-the-art SVM-based approach. Second, we addressed the complex clinical scenario of classifying PD and atypical parkinsonism. In this experiment, we showed the capability of a *motor tree* to discriminate between PD, DLB, PSP, and MSA.

## 2 Results

### 2.1 Lifespan tree of brain anatomy

Figure 1 provides a conceptual overview of the proposed *lifespan tree* used to model the progression of brain anatomy in neurodegenerative diseases. Lifespan tree combines lifespan modelling[19] and non-linear dimensionality reduction[25] yielding a compact representation of brain anatomy evolution across different pathologies over the entire lifespan. First, for each cerebral structure (n=124) and each considered population, a volume trajectory with respect to age is estimated across the entire lifespan to obtain normal and pathological lifespan models. Second, the multiple lifespan trajectories are reduced using a manifold learning method (i.e., UMAP). Finally, the 3D lifespan tree of brain anatomy is reconstructed by using age as the vertical axis. This provide a tree-like structure with a common trunk at early ages follow by progressive divergences related to differential patterns of atrophy, each summarized as a specific location within the reduced space. This novel tool offers a visually interpretable, rapid and flexible framework for the differential diagnosis of various neurological disorders. The position of an external test subject within the 3D tree space directly suggests a potential diagnosis and a final score for each diagnosis is numerically provided to the clinician.

### 2.2 Cognitive tree for the differential diagnosis of cognitive neurodegenerative diseases

To demonstrate the differential diagnosis capability of the *lifespan tree* of brain anatomy, we first tested a challenging classification task distinguishing between normal aging and six cognitive disorders. For this purpose, we used brain MRI from 461 patients with AD, 155 with bvFTD, 39 with semantic dementia, 41 with PNFA, 91 with PSP, and 126 with DLB for the construction of the branches of a *cognitive tree*, as well as 5,346 healthy young subjects to build the trunk of the tree and 31,181 cognitively normal subjects older than 44y to build the cognitively normal (CN) branch (see Table 1 and supp Table 1 for details).

For testing, we used an independent dataset consisting of 528 CN, 488 AD, 90 bvFTD, 44 semantic dementia, 40 PNFA, 67 PSP, and 47 DLB subjects from entirely independent external databases (see Table 1 and supp Table 2 for details).

Table 2 presents the classification results obtained on the external testing dataset in terms of Balanced accuracy (BACC) and sensitivity (SEN, or Recall: the proportion of correct positive predictions out of all positive predictions) per class. Instead of Area

Under the Curve, BACC is preferred since it is a more appropriate measure for multiclass classification when observations are not balanced between each class. For each metric, we estimated the top-1, top-2 and top-3 values, and their corresponding confidence intervals. The top-1 metrics is the usual metric (i.e., the top prediction is the correct class). The top-2 metric measures if the correct class is in the top 2 predictions and top-3 if the correct class is in the top 3 predictions. Confidence intervals were estimated using bootstrap with 95% confidence (N=10000) and p-value between BACC with a paired hypothesis testing[26]. In this experiment, the *lifespan tree* of brain anatomy obtained significantly better top-1 BACC (p-value < 0.0001) with 55% (randomness being 14%) while the SVM-based method achieved 46% top-1 BACC, indicating a lack of sensitivity for some classes, such as DLB, and an over-detection of CN class (see Table 2 and confusion matrices in supplementary Figure 1). As expected, all the top-2 metrics increased, and our lifespan tree reached a significantly better top-2 BACC (p-value=0.0025) of 71% (randomness being 29%), while the SVM-based method achieved a top-2 BACC of 65%. Finally, for the top-3 metric (randomness being 43%), both methods obtained a similar BACC around 79% (p-value=0.45). This demonstrates that SVM is similar to our method to rank the correct answer within their top 3 predictions out of seven but our method is significantly better in accurately assigning the correct class in the first or second position.

In addition to differential diagnosis metrics, we evaluated the ability of the methods to detect abnormality (i.e., patients vs. controls) by combining the pathological populations (AD, DLB, PSP, bvFTD, PNFA, SD) into a single class before computing the metrics. In this experiment, the *lifespan tree* obtained significantly better BACC (p-value < 0.03) with 78% (randomness being 50%) while the SVM-based method achieved 76% BACC (see Table 3). Moreover, *lifespan tree* was able to detect cognitive disorders with a sensitivity of 83%, while the SVM-based method had a much lower sensitivity of 54%, missing a significant number of cases.

Figure 2 presents our *cognitive tree* encompassing normal aging and six cognitive disorders (resulting in seven branches). The tree is cut at age 60y to better visualize the distances between branches at this age, as an illustrative example. We can observe that the branch closest to CN is DLB. Additionally, the DLB branch is situated between the CN and AD branches. As a result, some AD patients (44% top-1 sensitivity) were misclassified as CN, some CN subjects were identified as DLB patients, and many DLB patients (with only 32% top-1 sensitivity) were incorrectly assigned to the AD and CN classes (see Table 2 and confusion matrices in supplementary Figure 1).

The proximity of the PSP and PNFA branches indicates that these disorders share anatomical similarities. PSP was well-detected with a top-1 sensitivity of 71%, due to its specific atrophy pattern focused on the thalamus, brainstem, and ventral diencephalon (see Figure 3). Conversely, PNFA yielded a low top-1 sensitivity of 33%, as each part of its atrophic pattern could be found in at least one other population (see Figure 3).

Similarly, distinct branches were observed for semantic dementia and bvFTD. Semantic dementia subjects were easily distinguishable with 91% top-1 sensitivity, thanks to a highly pronounced atrophy pattern (see Figure 3). In contrast, bvFTD patients were more challenging to classify, with a top-1 sensitivity of 43%. While on average this population is distinct from others (i.e., in terms of distance between branches), the classification of test subjects exhibits high variability (see confusion matrices in supplementary Figure 1). This might be due to the anatomical heterogeneity of the bvFTD population.

## 2.3 Motor tree for differential diagnosis of PD and atypical parkinsonism

To further demonstrate the clinical relevance of our method, we performed another classification task for distinguishing between disorders that share parkinsonian symptoms, where the differential diagnosis is particularly challenging. For *motor tree* construction, we used brain MRI from 5,213 healthy subjects younger than 40y for the trunk of the tree and 91 PSP, 126 DLB, 133 PD, and 21 MSA subjects for the branches. For testing, we used 67 PSP, 47 DLB, 333 PD, and 117 MSA subjects from independent datasets (see Table 1).

Table 4 presents the classification results on the external datasets. Similar to the previous experiment, *the lifespan tree* of brain anatomy provided a significantly better (p-value = 0.03) top-1 BACC at 62% (with randomness being 25%) compared to the SVM-based method (56% top-1 BACC). Furthermore, for the top-2 metrics (with randomness being 50%), both methods obtained a similar BACC of around 80% (p-value = 0.13). This shows again that our method is better than SVM in detecting the correct class but both methods similarly rank the correct answer within their top-2 predictions out of four.

As previously, we also evaluated the ability of the methods to detect atypical parkinsonism vs. PD by combining the atypical populations (DLB, PSP and MSA) into a single class before computing the metrics. In this experiment, both methods obtained similar BACC (p-value = 0.11) with 84% (randomness being 50%) for lifespan tree and with 86% for the SVM-based method achieved (see Table 5). Moreover, both methods reached similar sensitivities, with 87% for *lifespan tree* and 86% for SVM-based method.

Figure 4 depicts the *motor tree* with the four branches representing PD and three causes of parkinsonism. The cut tree illustrates the proximity of DLB and PD branches at 60 years old, PD being characterized by the lowest atrophic pattern within the population compared to normal aging (see Figure 5). In contrast, PSP and MSA exhibited more distinct anatomical patterns. However, both methods, especially SVM, encountered difficulties in distinguishing MSA patients from PD patients (see confusion matrices in supplementary Figure 2), resulting in low top-1 sensitivity for MSA (ranging from 27% for the SVM-based method to 42% for the *lifespan tree* of brain anatomy) although our lifespan tree produced higher sensitivity for this population. Similar to the *cognitive tree* experiment, all methods achieved their highest sensitivity on the PSP population, which possesses a clear and specific MRI phenotype compared to other classes (see Figure 5).

## 3  Discussion

So far, the large majority of AI-enhanced MRI analysis has been proposed for binary classification tasks between a specific pathology and healthy conditions (see supplementary tables 6 and 7 for validation of the *cognitive and motor trees* on binary tasks) and mainly on AD diagnosis and prognosis (see supplementary tables 3-5 and supplementary Figure 3 for validation of our method on AD tasks). However, this over-simplification of diagnostic decision-making does not reflect everyday clinical practice. In this paper, we introduced a novel framework called *lifespan tree* that we used here to model the progression of brain anatomy in neurodegenerative disorders. This framework offers an elegant representation of the anatomical changes associated with neurodegeneration. We conducted two experiments demonstrating the value of this method for both motor and cognitive disorders and demonstrated that *lifespan tree* enables efficient and interpretable differential diagnosis between multiple disorders simultaneously. Additionally, the tree's shapes can provide new insights into the MRI phenotypes' similarities among different disorders. Moreover, as demonstrated in previous studies[19–23], lifespan modeling allows to measure the distances (i.e., the severity of atrophy progression) between normal aging and different pathological models to be visualized on an atlas (see Figures 3 and 5). This enables easy visualization of the archetypal progression of each disorder according to age which can be compared to the anatomy of each patient (see cut tree in Figure 1). Throughout our experiments, we have demonstrated that *cognitive* and *motor trees* consistently outperformed state-of-the-art SVM-based methods to discriminate cognitive and motor neurodegenerative diseases.

First, we applied our framework for the differential diagnosis of seven diseases or syndromes by constructing a *cognitive tree*. We demonstrated that this seven-branch tree was capable of distinguishing CN, AD, DLB, bvFTD, PNFA, semantic dementia and PSP with a top-1 BACC of 55% (randomness being 14%) on completely external datasets, compared to 46% for SVM on volumes, which is considered as the state of the art[17]. Moreover, our *lifespan tree* was able to detect that a subject has a cognitive disorder (i.e., controls vs. patients) with 83% of sensitivity compared to only 54% for the SVM-based method.

While there is no directly comparable experiment in the literature, previous studies have proposed methods for the differential diagnosis of various types of cognitive disorders. For the 3-class classification problem of CN vs. AD vs. FTD, SVM on volumes achieved top-1 BACCs from 75% to 80%[12,14], while recent end-to-end deep learning methods achieved BACC from 67% to 83%[27,28]. These results suggest that machine learning methods remain competitive compared with deep learning methods for such differential diagnosis task. In 5-class classification problems, machine learning on brain volumes achieved top-1 BACCs from 51%[29] to 66%[30] for CN vs. AD vs. FTD vs. DLB vs. vascular dementia, and 59% for subjective memory complaint vs. AD vs. FTD vs. DLB vs. vascular dementia[16]. More recently, a 7-class method dedicated to comparing AD vs. bvFTD vs. semantic dementia vs. PNFA vs. PSP vs. corticobasal syndrome vs. logopenic variant of primary progressive aphasias has been proposed[17]. For this 7-class differential diagnosis task, SVM on volumes obtained 47% of top-1 BACC which is similar to the 46% presented here using SVM on volumes on a slightly different 7-class differential diagnosis problem. This comparison highlights the good performance (55% of top-1 BACC) of our *lifespan tree* of brain anatomy for 7-class differential diagnosis in the field of cognitive neurology. Furthermore, it is crucial to note that unlike all the mentioned methods, which were validated using cross-validation on the same databases (potentially leading to overoptimistic performances[31]), our method was validated on completely external datasets, highlighting its robustness in clinical real-world scenarios.

During our experiments with the *cognitive tree*, our method achieved top-1 sensitivities of 44% for AD, 74% for CN, 32% for DLB, 43% for bvFTD, 33% for PNFA, 91% for semantic dementia, and 71% for PSP (see Table 2). The populations with the lowest sensitivity were DLB and PNFA. For DLB, our results were consistent with[16], where DLB achieved the lowest sensitivity (38%) compared to other populations (subjective memory complaint, AD, FTD and vascular dementia) within their 5-class problem. This low sensitivity may arise from the proximity of MRI phenotypes between AD and DLB (see Figure 3), the mild severity of brain atrophy usually reported in DLB (see Figure 3)[32,33], the fact that a significant number of DLB patients might have AD as co-pathology[34] and the heterogeneity of the DLB population[35]. Regarding PNFA, our results are similar to those of[17] where the authors obtained a sensitivity of 36% in their 7-class problem. This is probably explained by the heterogeneity of the underlying neuropathology explaining this syndromic presentation of progressive aphasia, that may overlap distinct tauopathies including PSP for instance[36]. In their study, they also demonstrated that semantic dementia is the easiest variant of FTD to classify, a finding confirmed in our results (see Table 2). Furthermore, they obtained similar top-1 sensitivity results to ours for AD (42%). These comparisons on cognitive disorders highlight the consistency of our results with existing literature and further validate the enhanced effectiveness of our lifespan tree model of brain anatomy for differential diagnosis among multiple cognitive neurodegenerative diseases, compared to state-of-the-art SVM methods.

Second, we proposed a 4-class differential diagnosis approach for atypical parkinsonism by designing a *motor tree*. This experiment aimed to demonstrate the capability of our *lifespan tree* of brain anatomy method to discriminate between PD, DLB, PSP and MSA, illustrating the potential of this novel framework to be applied across a broad range of scenarios. In this experiment, our *motor tree* achieved a top-1 BACC of 62% (randomness being 25%), while SVM on brain volumes attained 56% of top-1 BACC on completely external databases (see Table 3). Moreover, *lifespan tree* and SVM reached similar sensitivities (around 86%) to detect that a subject has an atypical parkinsonism (i.e., PD vs. PSP or MSA or DLB). To date, there are no directly comparable experiments in the literature. However, for the easiest 3-class classification problem of CN vs. PD vs. PSP, SVM on volumes has previously achieved a top-1 BACC of 67% using cross-validation[10].

During our experiment, the *motor tree* achieved top-1 sensitivities of 66% for DLB, 58% for PD, 81% for PSP, and 42% for MSA. Of note, the MSA diagnostic criteria[37] used in the training dataset were not the most recent, and the clinical details available in the database did not allow for the reconstruction of the current criteria proposed by the *Movement Disorder Society*[4], which distinguishes between MSA-P (parkinsonian type) and MSA-C (cerebellar type). Consequently, a mixture of MSA-P and MSA-C cases may have been used for training (see divergence of cerebellum compared to normal aging in MSA model on Figure 5), unlike the test dataset where only patients diagnosed with MSA-P. The lower sensitivity for MSA, which is a rare disease, might also be explained by the sample size of the training dataset (only 21 patients).

In this study, we exclusively utilized volumes extracted from structural MRI as features. However, it is well recognized that incorporating additional variables, such as those derived from other imaging modalities, clinical scores, or fluid biomarkers, has the potential to significantly enhance classification results[10,13,15,16,29,30]. The *lifespan tree* of brain anatomy method can readily accommodate such additional variables. Nevertheless, one of the main challenges lies in finding datasets that provide consistent imaging modalities, clinical scores, and fluid biomarkers for all subjects involved. Given that T1-weighted (T1w) MRI data is widely available in clinical setting and across the datasets used in this study, we chose to first focus on this classical feature type in this work. The extension of our *motor* and *cognitive trees* to incorporate multimodal imaging and clinical scores will be a key focus of our future works. By integrating diverse data sources, we aim to further enhance the accuracy and robustness of our differential diagnosis framework. The implementation of our method in clinical practice could provide an anatomical score of belonging to a class (i.e., see Figure 1) that the clinician could compare with the demographic, clinical and biological data in their possession.

Finally, transitioning from binary (as usually done in the literature) to multiclass classification presents a considerable increase in complexity. Binary classification tasks, such as distinguishing patients from controls, yield higher performance measures due to their relatively simpler nature (see results of binary classification achieved with our framework in supplementary Tables 3-7). However, the clinical reality demands a more nuanced approach that extends beyond binary decisions. In clinical practice, neurologists and radiologists deal with complex and/or diffuse atrophy patterns providing little information. Moreover, patterns of atrophy that easily differentiate AD from healthy subjects (see supplementary Table 4) become less informative when considering differential diagnoses (see Table 1). This is typically the case with hippocampal atrophy, which can be encountered in many neurodegenerative diseases[21,22,38]. Therefore, to achieve accurate differential diagnoses, it becomes imperative to conduct a comprehensive analysis of multiple brain regions and considering the nuanced interplay of various pathologies. This is where AI might play a crucial role. By systematically analyzing the entire brain and integrating information from multiple regions, these algorithms can assist neurologists and radiologists in identifying the most suitable diagnosis based on the overall atrophy pattern. *The lifespan tree* of brain anatomy serves as both a means of analysis and a summary of this information, making it easily understandable for clinicians. Specifically, our method can embed volumetric data from 124 brain structures into a latent space where participants with similar atrophy topographies cluster together, while those with differing atrophy patterns are positioned further apart. This enables the summarization of complex atrophy patterns.

# 4 Methods

## 4.1 Dataset description

Training dataset (N=37594) for the construction of lifespan tree of brain anatomy

Our training dataset was composed of 37,594 baseline T1w MRI from 20 open access databases: ABIDE[39], ADHD200[40], ADNI[41], ALLFTD, AOMIC[42], Calgary[43], CamCAN[44], C-MIND, DLBS, ICBM[45], ISYB[46], IXI, MIRIAD[47], NDAR[48], OASIS[49], PDBP[50], PIXAR, SALD[51], SLIM and UKBiobank[52] (see Table 1 and supplementary Table 1 for details). This dataset was composed of 36527 cognitively normal subjects (CN), 461 patients with AD, 126 patients with DLB, 155 patients with bvFTD, 41 with PNFA, 39 with semantic dementia, 21 patients with MSA, 133 with PD and 91 with PSP.

Testing dataset (N=1754)

To validate our model, we built a testing dataset based on 6 external and independent databases: 4RTNI, AIBL[53], FMSA[54], NACC[55], NIFD[4], PPMI[56] (see Table 2). Therefore, we validated the generalization capacity of our method and its robustness to domain shift. This dataset was composed of 528 CN, 488 AD, 47 DLB, 90 BvFTD, 40 PNFA, 44 SD, 117 MSA, 333 PD and 67 PSP. For MSA, we selected only the subtype with predominant parkinsonism (MSA-P) in the FMSA cohort, i.e., we excluded subjects with predominant cerebellar dysfunction (MSA-C) who are not difficult to distinguish from PD and other parkinsonian disorders. For PD, we used only the sporadic form in the PPMI database, so we excluded all genetic forms.

## 4.2 MRI processing

All the images considered in our study underwent preprocessing using the AssemblyNet software (https://github.com/volBrain/AssemblyNet)[57]. AssemblyNet leverages collective artificial intelligence to achieve fine-grained segmentation (132 structures) of the entire brain in just 15 minutes.

The AssemblyNet preprocessing pipeline consisted of several key steps:

- Image denoising[58] to reduce noise and enhance image quality
- Inhomogeneity correction[59] to address intensity variations across the image
- Affine registration[60] to the Montreal Neurological Institute (MNI) space to ensure spatial alignment across subjects
- A second inhomogeneity correction in the MNI space[61] to further refine intensity uniformity.
- A final intensity standardization step[62] to ensure consistency in image intensity across subjects.

By systematically applying these preprocessing steps, we aimed to enhance the quality and consistency of the imaging data which is crucial to model the pseudo longitudinal course of atrophy by merging different patients and for comparisons of diseases from different databases.

After preprocessing, the brain underwent segmentation into several structures using an assembly of 250 deep learning models, as detailed in[57]. The segmentation process was based on the Neuromorphometrics protocol, which encompasses 132 structures[63]. This protocol follows the "general segmentation protocol" for subcortical structures, as defined by the "MGH Center for Morphometric Analysis" (http://neuromorphometrics.com/Seg/). Additionally, the segmentation of cortical structures adheres to the "BrainCOLOR protocol" (http://neuromorphometrics.com/ParcellationProtocol_2010-04-05.PDF). In our study, we focused on 124 structures, including all gray matter structures and the lateral ventricles. We removed the white matter structures less sensitive to neurodegenerative process.

Subsequently, we conducted a quality control (QC) procedure to meticulously select subjects for inclusion in our training dataset. For all training subjects flagged as failures by an AI-based automatic QC tool[64], a visual assessment was carried out. This involved individually inspecting the input images and segmentations generated by AssemblyNet using a 3D viewer. If the failure was confirmed by our expert, the subject was excluded from the training dataset (overall less than 2% of the training subjects were rejected). This rigorous QC process ensured the reliability and accuracy of the training data used in our study.

## 4.3 Volume normalization

To mitigate inter-subject variability, we employed normalization of all structure volumes using the intracranial cavity volume (ICV)[65]. Furthermore, we conducted z-score normalization of all normalized volumes, expressed as a percentage of ICV. To achieve this, we first calculated the mean and standard deviation for each structure using all CN subjects across the entire lifespan. Subsequently, for a given structure, we applied the same z-score normalization to all subjects, including those with pathologies.

Additionally, we addressed sex differences in volumes by employing a linear model when a significant sex effect was detected ($p < 0.05$), following the approach outlined in[20]. This ensured that volumes were appropriately compensated for sex differences. Consequently, by expressing volumes as z-scores of normalized and sex-corrected volumes, we effectively compensated for inter-subject, sex, and inter-structure variabilities. This normalization approach enhanced the comparability and robustness of our volumetric measurements across subjects and structures.

## 4.4 Lifespan brain volumetric trajectories estimation

To construct our lifespan models of brain structure volumes, we estimated normal and pathological trajectories of structure volumes across the entire lifespan, as illustrated in Figure 1. For each brain structure, separate models were estimated for each group to generate trajectories.

For the trajectories of CN, we used a large dataset comprising N=36527 subjects ranging from 1 year to 94 years of age, extracted from the training dataset.

In contrast, for the pathological trajectories, we followed the framework proposed in previous studies[19–24] which involves a mixture of patients with young CN individuals. This framework operates under the assumption that neurodegeneration is a slow and progressive process meaning that the patients should have a normal volumetric pattern before the beginning of the disease (the trunk of the tree) and then to progressively diverge (the branches of the tree). Specifically, we incorporated all CN individuals younger than a certain age threshold. As outlined and justified in[20], we determined this threshold as the first centile value of the distribution of patients' ages in the training pathological populations, corresponding to 44 years for the dataset of patients with cognitive neurodegenerative diseases and 40 years old for the dataset of patients with PD and parkinsonism. Consequently, we used 5346 CN individuals younger than 44 years to build the *cognitive tree*, and utilized 5213 CN individuals younger than 40 years for the *motor tree*. This adaptive approach ensures that the threshold value is automatically adjusted based on the age distributions of the respective populations under consideration.

To estimate the volume trajectories, we evaluated several low-order polynomial models, including linear, quadratic, and cubic models. Following the approach outlined in previous studies[19–24]. A polynomial model was considered as a potential candidate only if it satisfied two criteria simultaneously:

- The F-statistic based on ANOVA, comparing the polynomial model to a constant model, was found to be significant ($p < 0.05$).
- All coefficients of the polynomial model were individually significant according to the T-statistic ($p < 0.05$).

Subsequently, to select the most relevant model among these potential candidates, we employed the Bayesian Information Criterion (BIC)[66]. The BIC facilitates model selection by balancing model fit and complexity, thereby aiding in the identification of the most parsimonious and informative model for describing the volume trajectories across different age and populations.

## 4.5 Lifespan tree construction

Herein, we introduce a novel lifespan method capable of automatically extracting pertinent information that effectively discriminates between groups of patients. This characteristic empowers our method to be applicable to any classification problem without requiring manual tuning. More significantly, our method is well-adapted in handling multi-class problems, making it particularly valuable for differential diagnosis tasks. Finally, our aim was to provide an interpretable tool by clinicians, therefore we decided to create a 3D representation of the brain progression over lifespan (higher dimension could be also used to better capture brain atrophy but making visualization harder).

To construct our 3D *lifespan tree* of the brain anatomy, the first step involves reducing the dimensionality of our feature space, which encompasses 125 features comprising 124 structure volumes and age, into a 3D space comprising 2D spatial coordinates along with age. For this purpose, we opted for a manifold learning approach leveraging UMAP (Uniform Manifold Approximation and Projection)[25]. UMAP is a nonlinear dimensionality reduction technique known for its ability to effectively separate patterns within high-dimensional datasets.

Addressing the challenge of incorporating the temporal nature of the data, we employed a 2D UMAP-based dimensionality reduction technique in conjunction with a posterior 3D reconstruction of the *lifespan tree* of brain anatomy. Additionally, to ensure effective separation of all considered populations (i.e., branches in the tree), including smaller ones, we utilized Monte Carlo simulation to generate synthetic samples. This approach enables us to enhance the discriminative capabilities of our *lifespan tree* of brain anatomy, ensuring clear separation between groups, as illustrated in Figure 1.

Recently, an approach called aligned-UMAP based on a sliding window framework has been introduced to handle time series, demonstrating its utility in analyzing longitudinal data[67]. While this method offers advantages for studying timeseries, its use can complicate the projection of new points into the reduced space due to the creation of different embeddings for each timepoint.

Given our objective of performing classification on external samples, we devised an alternative strategy centered on creating a unique unified 2D embedding and subsequently reconstructing the time axis (age axis in our case). Initially, we employed UMAP to reduce the dimensionality of all selected samples simultaneously from 124 dimensions (representing 124 structures) to a 2D space. Subsequently, we utilized the age information of the samples to project them into a 3D space, where the third axis corresponds to age (as depicted in Figure 1). Following this process, the 3D branches, representing the projection of the lifespan trajectories, can be plotted within our 3D space to construct the *lifespan tree*.

## 4.6 Synthetic sampling

To ensure the effectiveness of our global 2D embedding, careful selection of samples is crucial to achieve a robust separation of populations and maintain balanced representation across all populations and along ages. Instead of directly utilizing volumes (which may lack balance between populations and along ages), we propose employing synthetic samples generated through simulation based on lifespan trajectories. For each year, we generated 100 synthetic samples randomly drawn from a Gaussian distribution, with the lifespan trajectory values at that age serving as the mean and a standard deviation equal to 1 (given that trajectories were based on z-scores). All synthetic samples were projected to illustrate their distribution (branch size) within the final space, as depicted in Figure 1.

## 4.7 UMAP training

During UMAP training, we used the Matlab implementation of UMAP [68] with default parameters, excepted for the number of neighbors which was set to 10 times the number of considered populations. By this way, the number of neighbors is automatically adapted to the number of classes. Moreover, to enhance population separation, as done for trajectory estimation, we only used synthetic samples older than the first centile of patients' age in the training population (i.e., 44y for the cognitive tree and 40y for the motor tree). This ensures training the UMAP embedding over the most discriminative period of life.

## 4.8 Classification using *lifespan tree*

Once the 3D tree was built with the training dataset, we used it to classify subjects from the testing dataset (see Figure 1). To classify each subject, we projected them into the 3D tree space and identified the closest branch in terms of Euclidean distance, assigning the subject to the corresponding class. Additionally, we calculated the class membership score by using the squared distance within a Gaussian distribution with a mean of zero and a standard deviation of one. This score was then normalized by the sum of scores across all classes to provide a likelihood measure.

## 4.9 Classification using Support Vector Machine (SVM)

For differential diagnosis based on MRI biomarkers, Support Vector Machine (SVM) classification of volumes has demonstrated competitive results in recent literature[9–11,14]. Therefore, we used a multiclass SVM on volumes (124 features per subject) as our state-of-the-art reference to challenge the relevance of *the lifespan tree*. We also tested a variant incorporating volumes and age, but the results were slightly worse. The SVM-based method used the Matlab version of the multiclass error-correcting output codes classifier based on SVM learners with default parameters (https://www.mathworks.com/help/stats/classificationecoc.html).

## Data Availability statement

All the used MRI were from open access database and thus can be downloaded from the database provider website. AssemblyNet is freely available at https://github.com/volBrain/AssemblyNet. The proposed method will be made available through www.volbrain.net

## Statement of Contribution

P.C. developed the idea, the theoretical formalism, performed the analytic calculations and performed the numerical experiments. P.C., V.P. conceived and planned the experiments. P.P. and B.M. prepared and processed the data. P.C. took the lead in writing the manuscript. W.G.M., T.T., J. M. and V.P. aided in interpreting the results and worked on the manuscript. All authors provided critical feedback and helped shape the research, analysis and manuscript. All authors discussed the results and contributed to the final manuscript.

## Acknowledgements

This work benefited from the support of the project HoliBrain of the French National Research Agency (ANR-23-CE45-0020-01). Moreover, this project is supported by the Precision and global vascular brain health institute funded by the France 2030 investment plan as part of the IHU3 initiative (ANR-23-IAHU-0001). Finally, this study received financial support from the French government in the framework of the University of Bordeaux's France 2030 program / RRI "IMPACT and the PEPR StratifyAging. This work was also granted access to the HPC resources of IDRIS under the allocation 2022-AD011013848R1 made by GENCI. This work also benefited from the support of the project PID2020-118608RB-I00 (AEI/10.13039/501100011033) of the Ministerio de Ciencia e Innovacion de España.

Moreover, this work is based on multiple samples. We wish to thank all investigators of these projects who collected these datasets and made them freely accessible.

The C-MIND data used in the preparation of this article were obtained from the C-MIND Data Repository (accessed in Feb 2015) created by the C-MIND study of Normal Brain Development. This is a multisite, longitudinal study of typically developing children from ages newborn through young adulthood conducted by Cincinnati Children's Hospital Medical Center and UCLA and supported by the National Institute of Child Health and Human Development (Contract #s HHSN275200900018C). A listing of the participating sites and a complete listing of the study investigators can be found at https://research.cchmc.org/c-mind.

The NDAR data used in the preparation of this manuscript were obtained from the NIH-supported National Database for Autism Research (NDAR). NDAR is a collaborative informatics system created by the National Institutes of Health to provide a national resource to support and accelerate research in autism. The NDAR dataset includes data from the NIH Pediatric MRI Data Repository created by the NIH MRI Study of Normal Brain Development. This is a multisite, longitudinal study of typically developing children from ages newborn through young adulthood conducted by the Brain Development Cooperative Group and supported by the National Institute of Child Health and Human Development, the National Institute on Drug Abuse, the National Institute of Mental Health, and the National Institute of Neurological Disorders and Stroke (Contract #s N01- HD02-3343, N01-MH9-0002, and N01-NS-9-2314, -2315, -2316, -2317, -2319 and -2320). A listing of the participating sites and a complete listing of the study investigators can be found at http://pediatricmri.nih.gov/nihpd/info/participating_centers.html

The ADNI data used in the preparation of this manuscript were obtained from the Alzheimer's Disease Neuroimaging Initiative (ADNI) (National Institutes of Health Grant U01 AG024904). The ADNI is funded by the National Institute on Aging and the National Institute of Biomedical Imaging and Bioengineering and through generous contributions from the following: Abbott, AstraZeneca AB, Bayer Schering Pharma AG, Bristol-Myers Squibb, Eisai Global Clinical Development, Elan Corporation, Genentech, GE Healthcare, GlaxoSmithKline, Innogenetics NV, Johnson & Johnson, Eli Lilly and Co., Medpace, Inc., Merck and Co., Inc., Novartis AG, Pfizer Inc., F. Hoffmann-La Roche, Schering-Plough, Synarc Inc., as well as nonprofit partners, the Alzheimer's Association and Alzheimer's Drug Discovery Foundation, with participation from the U.S. Food and Drug Administration. Private sector contributions to the ADNI are facilitated by the Foundation for the National Institutes of Health (www.fnih.org). The grantee organization is the Northern California Institute for Research and Education, and the study was coordinated by the Alzheimer's Disease Cooperative Study at the University of California, San Diego. ADNI data are disseminated by the Laboratory for NeuroImaging at the University of California, Los Angeles. This research was also supported by NIH grants P30AG010129, K01 AG030514 and the Dana Foundation.

The OASIS data used in the preparation of this manuscript were obtained from the OASIS project funded by grants P50 AG05681, P01 AG03991, R01 AG021910, P50 MH071616, U24 RR021382, R01 MH56584. See http://www.oasis-brains.org/ for more details.

The AIBL data used in the preparation of this manuscript were obtained from the AIBL study of ageing funded by the Common-wealth Scientific Industrial Research Organization (CSIRO; a publicly funded government research organization), Science Industry Endowment Fund, National Health and Medical Research Council of Australia (project grant 1011689), Alzheimer's Association, Alzheimer's Drug Discovery Foundation, and an anonymous foundation. See www.aibl.csiro.au for further details.

The ICBM data used in the preparation of this manuscript were supported by Human Brain Project grant PO1MHO52176-11 (ICBM, P.I. Dr John Mazziotta) and Canadian Institutes of Health Research grant MOP- 34996.

The IXI data used in the preparation of this manuscript were supported by the U.K. Engineering and Physical Sciences Research Council (EPSRC) GR/S21533/02 - http://www.brain-development.org/ .

The ABIDE data used in the preparation of this manuscript were supported by ABIDE funding resources listed at http://fcon_1000.projects.nitrc.org/indi/abide/. ABIDE primary support for the work by Adriana Di Martino was provided by the NIMH (K23MH087770) and the Leon Levy Foundation. Primary support for the work by Michael P. Milham and the INDI team was provided by gifts from Joseph P. Healy and the Stavros Niarchos Foundation to the Child Mind Institute, as well as by an NIMH award to MPM (R03MH096321). http://fcon_1000.projects.nitrc.org/indi/abide/

Data used in the preparation of this article were obtained from the MIRIAD database. The MIRIAD investigators did not participate in analysis or writing of this report. The MIRIAD dataset is made available through the support of the UK Alzheimer's Society (Grant RF116). The original data collection was funded through an unrestricted educational grant from GlaxoSmithKline (Grant 6GKC).

The ADHAD, DLBS and SALD data used in the preparation of this article were obtained from http://fcon_1000.projects.nitrc.org (Mennes et al., NeuroImage, 2013; Wei et al., bioRxiv 2017).

Data used in the preparation of this article were obtained from the Parkinson's Progression Markers Initiative (PPMI) database (www.ppmi-info.org). PPMI—a public-private partnership—was funded by The Michael J. Fox Foundation for Parkinson's Research and funding partners that can be found at https://www.ppmi-info.org/about-ppmi/who-we-are/study-sponsor.

Data collection and sharing for this project was provided by the Cambridge Centre for Ageing and Neuroscience (CamCAN, https://camcan-archive.mrc-cbu.cam.ac.uk/dataaccess/). CamCAN funding was provided by the UK Biotechnology and Biological Sciences Research Council (grant number BB/H008217/1), together with support from the UK Medical Research Council and University of Cambridge, UK.

The 4-repeat tauopathy neuroimaging initiative (4RTNI) and frontotemporal lobar degeneration neuroimaging initiative (FTLDNI) were funded through the National Institute of Aging, and started in 2010. The primary goals of FTLDNI were to identify neuroimaging modalities and methods of analysis for tracking frontotemporal lobar degeneration (FTLD) and to assess the value of imaging versus other biomarkers in diagnostic roles. The Principal Investigator of NIFD was Dr. Howard Rosen, MD at the University of California, San Francisco. The data are the result of collaborative efforts at three sites in North America. For up-to-date information on participation and protocol, please visit http://memory.ucsf.edu/research/studies/nifd. Data collection and sharing for this project was funded by the Frontotemporal Lobar Degeneration Neuroimaging Initiative (National Institutes of Health). The study is coordinated through the University of California, San Francisco, Memory and Aging Center. FTLDNI data are disseminated by the Laboratory for Neuro Imaging at the University of Southern California.

The NACC database was funded by NIA/NIH Grants listed at https://naccdata.org/publish-project/authors-checklist#acknowledgment.

This research has been conducted using the UK Biobank Resource under application number 80509. See https://www.ukbiobank.ac.uk for further details.

The Amsterdam open MRI collection AOMIC ID-1000/PIOP1/PIOP2 data used in the preparation of this article were obtained from https://nilab-uva.github.io/AOMIC.github.io/ (Snoek L et al., Scientific data, 2021).

The Calgary preschool MRI dataset was available at https://osf.io/axz5r/ and supported by University of Calgary and CIHR (IHD-134090 & MOP-136797).

The Pixar database and related fundings were available at https://openneuro.org/datasets/ds000228/versions/1.1.0 (Richardson H et al., Nat Commun, 2018).

Data and biospecimens used in preparation of this manuscript were obtained from the Parkinson's Disease Biomarkers Program (PDBP) Consortium, supported by the National Institute of Neurological Disorders and Stroke at the National Institutes of Health. Investigators include: Roger Albin, Roy Alcalay, Alberto Ascherio, Thomas Beach, Sarah Berman, Bradley Boeve, F. DuBois Bowman, Shu Chen, Alice Chen-Plotkin, William Dauer, Ted Dawson, Paula Desplats, Richard Dewey, Ray Dorsey, Jori Fleisher, Kirk Frey, Douglas Galasko, James Galvin, Dwight German, Lawrence Honig, Xuemei Huang, David Irwin, Kejal Kantarci, Anumantha Kanthasamy, Daniel Kaufer, James Leverenz, Carol Lippa, Irene Litvan, Oscar Lopez, Jian Ma, Lara Mangravite, Karen Marder, Laurie Orzelius, Vladislav Petyuk, Judith Potashkin, Liana Rosenthal, Rachel Saunders-Pullman, Clemens Scherzer, Michael Schwarzschild, Tanya Simuni, Andrew Singleton, David Standaert, Debby Tsuang, David Vaillancourt, David Walt, Andrew West, Cyrus Zabetian, Jing Zhang, and Wenquan Zou. The PDBP Investigators have not participated in reviewing the data analysis or content of the manuscript.

Data collection and dissemination of the data presented in this manuscript was supported by the ALLFTD Consortium (U19: AG063911, funded by the National Institute on Aging and the National Institute of Neurological Diseases and Stroke) and the former ARTFL & LEFFTDS Consortia (ARTFL: U54 NS092089, funded by the National Institute of Neurological Diseases and Stroke and National Center for Advancing Translational Sciences; LEFFTDS: U01 AG045390, funded by the National Institute on Aging and the National Institute of Neurological Diseases and Stroke). The authors acknowledge the invaluable contributions of the study participants and families as well as the assistance of the support staffs at each of the participating sites.

# Figure legends

Figure 1: Schematic overview of the proposed method based on our *cognitive tree* (AD=Alzheimer's disease. bvFTD=behavioral variant frontotemporal dementia. CN=cognitively normal. DLB=dementia with Lewy bodies. PNFA=progressive non fluent aphasia. PSP=progressive supranuclear palsy. SD=semantic dementia). First, the training step involves converting the lifespan volumetric models of the 124 considered brain structures into a 3D lifespan tree (from N=37594). Synthetic sampling strategy is then applied to balance samples between populations and make uniform sample distributions along age. The nonlinear dimension reduction into a 2D space (x, y) is performed using UMAP manifold learning. The x and y axes are the first two components of the UMAP dimension reduction. The 3D tree is then reconstructed using age as the z-coordinate. Second, during the testing step, a new external test subject is projected into the 3D space of the lifespan tree (purple diamond). To achieve this, the volumes of the 124 brain structures are projected into 2D with the estimated UMAP transform to obtain x and y coordinates. The subject's age is then used as the z-coordinate in the lifespan tree. Finally, the distances between the subject's point and the tree branches are used to estimate the scores for each class. The closest branch (i.e., the highest score) determines the final class of the subject under study. It is important to note that the results provided to the clinician for each patient pertain to both the score of the primary diagnosis and the scores of various differential diagnoses. In this example, an AD patient (in purple) who is 80 years old is projected onto our cognitive lifespan tree. The closest branch is the AD branch, followed by the DLB branch. A cut tree at 80 years old can also be displayed to facilitate the detection of the closest branch (i.e., the class of the subject under study). Moreover, the distance to the normal aging lifespan model overlaid on an MRI atlas can be displayed to assist clinicians in their decision-making.

Figure 2: Left: The 3D cognitive tree composed by seven branches for Cognitively Normal (CN), Alzheimer's Disease (AD), behavioral variant frontotemporal dementia (bvFTD), semantic dementia (SD), progressive non fluent aphasia (PNFA), progressive supranuclear palsy (PSP) and dementia with Lewy bodies (DLB). Right: A 2D tree cut at 60y (including test subjects between 58y and 62y). The cloud dots surrounding the branches are the synthetic samples resulting from the Monte Carlo simulation. The diamonds represent the test subjects.

Figure 3: Atlas of the progression of the distance to normal aging (i.e., atrophy severity) for each neurodegenerative diseases reported in the cognitive tree. The four columns represent the anatomical progression of each disease at 60y, 70y, 80y and 90y. AD=Alzheimer's Disease, bvFTD=behavioral variant frontotemporal dementia, DLB= dementia with Lewy bodies, SD=semantic dementia, PNFA=progressive non fluent aphasia, PSP=progressive supranuclear palsy.

Figure 4: Left: The 3D motor tree composed by four branches for Parkinson's Disease (PD), progressive supranuclear palsy (PSP), dementia with Lewy bodies (DLB) and multiple system atrophy (MSA). Right: A 2D tree cut at 60y (including test subjects between 58y and 62y). The cloud dots surrounding the branches are the synthetic samples resulting from the Monte Carlo simulation. The diamonds represent the test subjects.

Figure 5: Atlas of the progression of the distance to normal aging (i.e., atrophy severity) for each neurodegenerative diseases reported in the motor tree. The four columns represent the anatomical progression of each disease at 60y, 70y, 80y and 90y. DLB=dementia with Lewy bodies, MSA=multiple system atrophy, PD=Parkinson's Disease, PSP=progressive supranuclear palsy.

# Tables

*Table 1: Description of the databases used for the construction (training) and for the testing of the cognitive and motor trees. The table provides the number of subjects (N) per class, the sex distribution between female (F) and male (M) the mean age [age range]. Testing and training datasets come from totally independent databases (see supp Table 1 and supp Table 2 for details). During our experiments, we used brain MRI from cognitively normal subjects (CN), patients with Alzheimer's disease (AD), dementia with Lewy bodies (DLB), progressive supranuclear palsy (PSP), behavioral variant frontotemporal dementia (bvFTD), progressive non-fluent aphasia (PNFA) and semantic dementia (SD), Parkinson's disease (PD) and multiple system atrophy (MSA).*

|  | CN | AD | DLB | bvFTD | PNFA | SD | MSA | PD | PSP |
|---|---|---|---|---|---|---|---|---|---|
| Training N= 37594 | N = 36527<br>F:18090 / M:18437<br>58 [1 – 94] | N = 461<br>F:229 / M:232<br>74 [52 – 96] | N = 126<br>F:22 / M:104<br>69 [50 – 90] | N = 155<br>F:58 / M:97<br>64 [40 – 83] | N = 41<br>F:22 / M:19<br>69 [55 – 82] | N = 39<br>F:16 / M:23<br>64 [50 – 79] | N = 21<br>F:4 / M:17<br>65 [53 – 80] | N = 133<br>F:48 / M:85<br>67 [40 – 81] | N = 91<br>F:48 / M:43<br>68 [52 – 80] |
| Testing N= 1754 | N = 528<br>F:308 / M:220<br>70 [30 – 100] | N = 488<br>F:304 / M:184<br>74 [46 – 96] | N = 47<br>F:8 / M:39<br>74 [55 – 89] | N = 90<br>F:31 / M:59<br>62 [45 – 76] | N = 40<br>F:23 / M:17<br>69 [54 – 81] | N = 44<br>F:20 / M:24<br>64 [50 – 85] | N = 117<br>F:60 / M:57<br>66 [42 – 84] | N = 333<br>F:120 / M:213<br>62 [34 – 83] | N = 67<br>F:38 / M:29<br>71 [55 – 86] |

Table 2: Comparison between the seven-branch cognitive tree and the state-of-the-art SVM method for the differential diagnosis of cognitive disorders (using external validation datasets). BACC=balanced accuracy. SEN=sensitivity (recall). SPE=specificity. For each metric, the confidence interval is provided between brackets. AD=Alzheimer's disease. bvFTD=behavioral variant frontotemporal dementia. CN=cognitively normal. DLB=dementia with Lewy bodies. PNFA=progressive non fluent aphasia. PSP=progressive supranuclear palsy. SD=semantic dementia. SEN=sensitivity (recall). Top-1 means that the correct class is the prediction with the highest probability (usually measurements). Top-2 means that the correct class is within the predictions with the two highest probabilities. Top-3 means that the correct class is within the predictions with the three highest probabilities. The best results are in red. * indicates significantly better BACC with p-value < 0.05.

| Classification on external datasets (%) | BACC Top-1 | SEN. AD Top-1 | SEN. CN Top-1 | SEN. DLB Top-1 | SEN. bvFTD Top-1 | SEN. PNFA Top-1 | SEN. SD Top-1 | SEN. PSP Top-1 |
|---|---|---|---|---|---|---|---|---|
| **Lifespan tree** | 55* [52-59] | 44 [39-48] | 74 [70-78] | 32 [19-46] | 43 [33-54] | 33 [18-48] | 91 [81-98] | 71 [61-82] |
| **SVM on volumes** | 46 [42-49] | 43 [38-47] | 97 [96-98] | 0 [0-0] | 51 [41-61] | 17 [6-30] | 57 [41-72] | 58 [46-70] |
|  | Top-2 | Top-2 | Top-2 | Top-2 | Top-2 | Top-2 | Top-2 | Top-2 |
| **Lifespan tree** | 71* [68-75] | 60 [56-65] | 80 [77-83] | 72 [59-85] | 59 [48-69] | 50 [34-66] | 95 [88-100] | 82 [72-91] |
| **SVM on volumes** | 65 [61-69] | 77 [73-81] | 100 [99-100] | 13 [4-23] | 70 [60-79] | 43 [27-58] | 75 [61-87] | 79 [69-88] |
|  | Top-3 | Top-3 | Top-3 | Top-3 | Top-3 | Top-3 | Top-3 | Top-3 |
| **Lifespan tree** | 79 [76-83] | 72 [68-76] | 94 [92-96] | 89 [80-98] | 59 [49-69] | 57 [41-73] | 95 [88-100] | 85 [76-93] |
| **SVM on volumes** | 78 [75-82] | 89 [87-92] | 100 [100-100] | 47 [32-61] | 81 [73-89] | 55 [39-70] | 86 [75-96] | 91 [84-97] |

Table 3: Comparison of the seven-branch cognitive tree with the state-of-the-art SVM method in detecting the presence of pathology (i.e., 528 controls vs. 776 patients). For this analysis, the six pathological populations (AD, DLB, PSP, bvFTD, PNFA, SD) were merged into a single class. BACC=balanced accuracy. SEN=sensitivity. SPE=specificity. The best-performing results are highlighted in red. * denotes a significantly higher BACC with a p-value < 0.05.

| Classification on external datasets (%) | BACC | SEN. | SPE. |
|---|---|---|---|
| **Lifespan tree** | **78*** [76-81] | 83 [80-85] | 74 [70-78] |
| **SVM on volumes** | 76 [74-78] | 54 [51-56] | 97 [96-99] |

*Table 4: Comparison between the four-branch motor tree and the state-of-the-art SVM method for the differential diagnosis of Parkinson's Disease and parkinsonism (PSP, DLB, MSA) (using external validation datasets). BACC= balanced accuracy. For each metric, the confidence interval is provided between brackets DLB=dementia with Lewy bodies. MSA=multiple system atrophy. PD=Parkinson Disease. PSP=progressive supranuclear palsy. SEN=sensitivity (recall). Top-1 means that the correct class is the prediction with the highest probability. Top-2 means that the correct class is within the predictions with the two highest probabilities. The best results are in red. * indicates significantly better BACC with p-value < 0.05.*

| Classification on external datasets (%) | BACC Top-1 | SEN. DLB Top-1 | SEN. MSA Top-1 | SEN. PD Top-1 | SEN. PSP Top-1 |
|---|---|---|---|---|---|
| **Lifespan tree** | 62* [58-67] | 66 [52-79] | 42 [33-51] | 58 [52-62] | 81 [71-90] |
| **SVM on volumes** | 56 [52-60] | 45 [30-60] | 27 [19-35] | 68 [62-73] | 87 [78-94] |
| | Top-2 | Top-2 | Top-2 | Top-2 | Top-2 |
| **Lifespan tree** | 80 [76-84] | 86 [75-95] | 67 [58-75] | 81 [77-86] | 85 [79-94] |
| **SVM on volumes** | 81 [77-85] | 74 [61-86] | 66 [57-74] | 87 [83-90] | 97 [92-100] |

*Table 5: Comparison of the four-branch motor tree with the state-of-the-art SVM method in detecting the presence of atypical PD (i.e., 333 PD vs. 231 atypical PD). For this analysis, the three atypical populations (DLB, MSA, PSP) were merged into a single class. BACC=balanced accuracy. SEN=sensitivity. SPE=specificity. The best-performing results are highlighted in red. * denotes a significantly higher BACC with a p-value < 0.05.*

| Classification on external datasets (%) | BACC | SEN. | SPE. |
|---|---|---|---|
| **Lifespan tree** | 84 [79-89] | 87 [84-90] | 81 [71-90] |
| **SVM on volumes** | 86 [82-90] | 86 [83-89] | 86 [82-94] |

Lifespan models

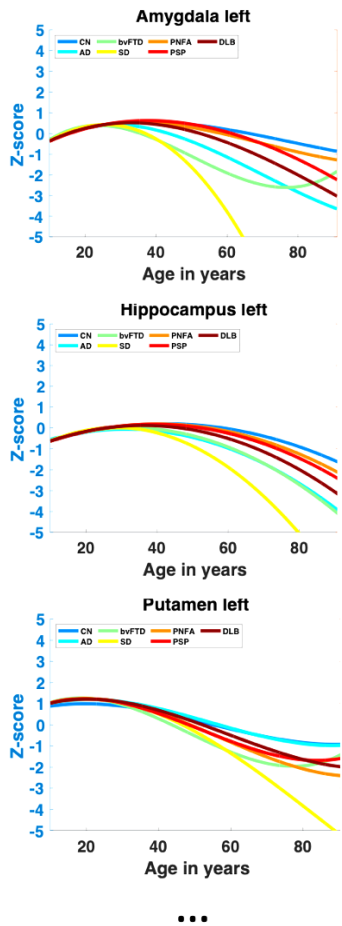

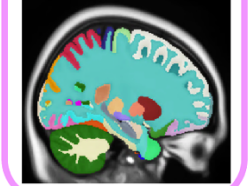
Segmentation

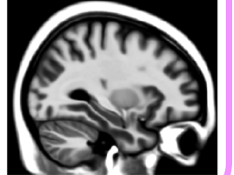
New Subject
80y old

3D Lifespan Tree
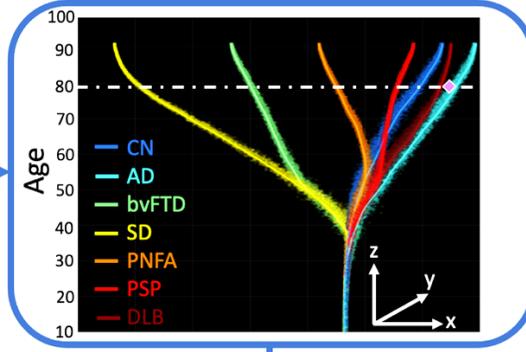

Scores
- CN 16.7%
- AD 38.5%
- bvFTD 0.8%
- SD 0.0%
- PNFA 2.7%
- PSP 7.9%
- DLB 33.4%

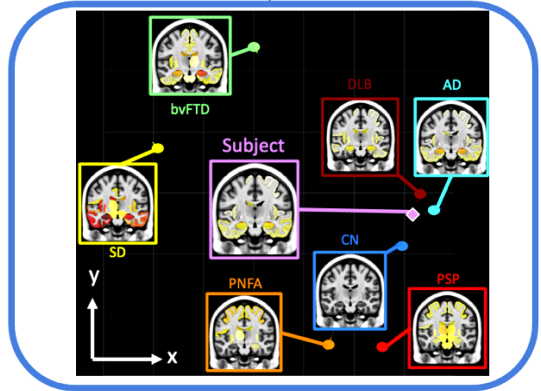
2D Cut Tree at 80y

*Figure 1: Schematic overview of the proposed method based on our cognitive tree (AD=Alzheimer's disease. bvFTD=behavioral variant frontotemporal dementia. CN=cognitively normal. DLB=dementia with Lewy bodies. PNFA=progressive non fluent aphasia. PSP=progressive supranuclear palsy. SD=semantic dementia). First, the training step involves converting the lifespan volumetric models of the 124 considered brain structures into a 3D lifespan tree (from N=37594). Synthetic sampling strategy is then applied to balance samples between populations and make uniform sample distributions along age. The nonlinear dimension reduction into a 2D space (x, y) is performed using UMAP manifold learning. The x and y axes are the first two components of the UMAP dimension reduction. The 3D tree is then reconstructed using age as the z-coordinate. Second, during the testing step, a new external test subject is projected into the 3D space of the lifespan tree (purple diamond). To achieve this, the volumes of the 124 brain structures are projected into 2D with the estimated UMAP transform to obtain x and y coordinates. The subject's age is then used as the z-coordinate in the lifespan tree. Finally, the distances between the subject's point and the tree branches are used to estimate the scores for each class. The closest branch (i.e., the highest score) determines the final class of the subject under study. It is important to note that the results provided to the clinician for each patient pertain to both the score of the primary diagnosis and the scores of various differential diagnoses. In this example, an AD patient (in purple) who is 80 years old is projected onto our cognitive lifespan tree. The closest branch is the AD branch, followed by the DLB branch. A cut tree at 80 years old can also be displayed to facilitate the detection of the closest branch (i.e., the class of the subject under study). Moreover, the distance to the normal aging lifespan model overlaid on an MRI atlas can be displayed to assist clinicians in their decision-making.*

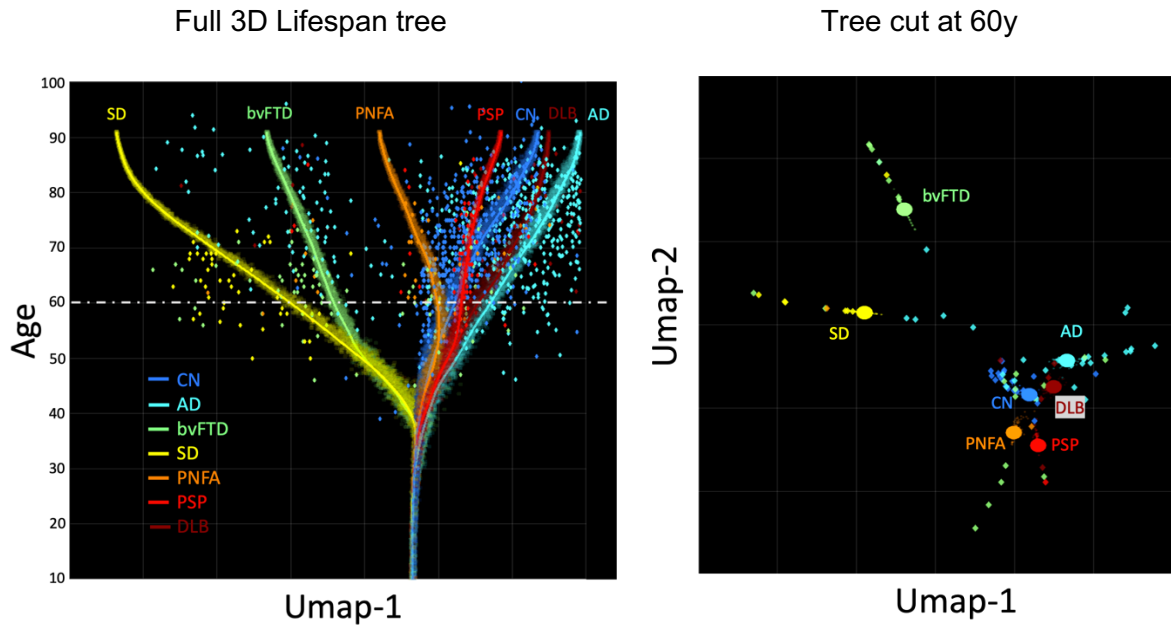

*Figure 2: Left: The 3D cognitive tree composed by seven branches for Cognitively Normal (CN), Alzheimer's Disease (AD), behavioral variant frontotemporal dementia (bvFTD), semantic dementia (SD), progressive non fluent aphasia (PNFA), progressive supranuclear palsy (PSP) and dementia with Lewy bodies (DLB). Right: A 2D tree cut at 60y (including test subjects between 58y and 62y). The cloud dots surrounding the branches are the synthetic samples resulting from the Monte Carlo simulation. The diamonds represent the test subjects.*

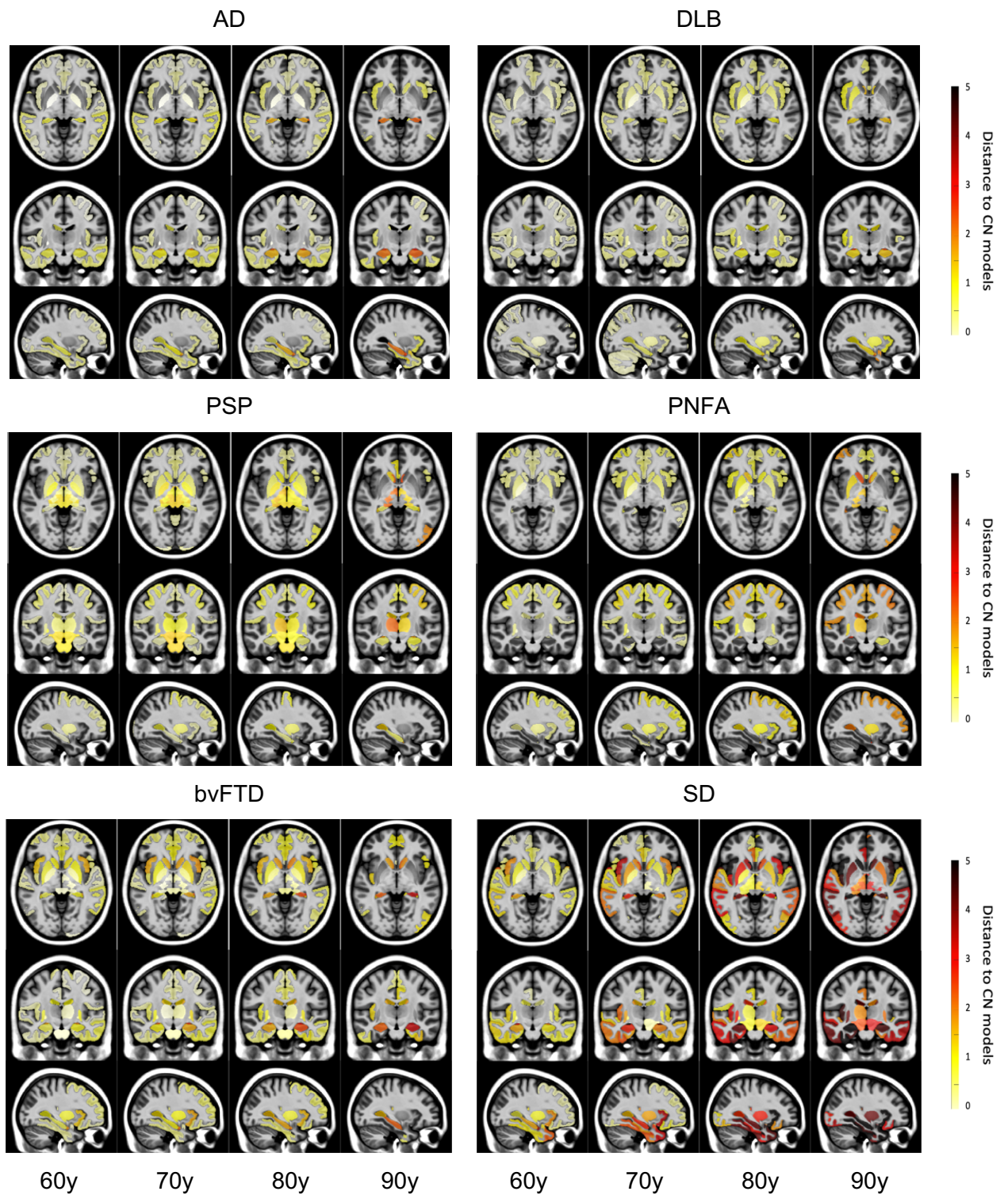

*Figure 3: Atlas of the progression of the distance to normal aging (i.e., atrophy severity) for each neurodegenerative diseases reported in the cognitive tree. The four columns represent the anatomical progression of each disease at 60y, 70y, 80y and 90y. AD=Alzheimer's Disease, bvFTD=behavioral variant frontotemporal dementia, DLB= dementia with Lewy bodies, SD=semantic dementia, PNFA=progressive non fluent aphasia, PSP=progressive supranuclear palsy.*

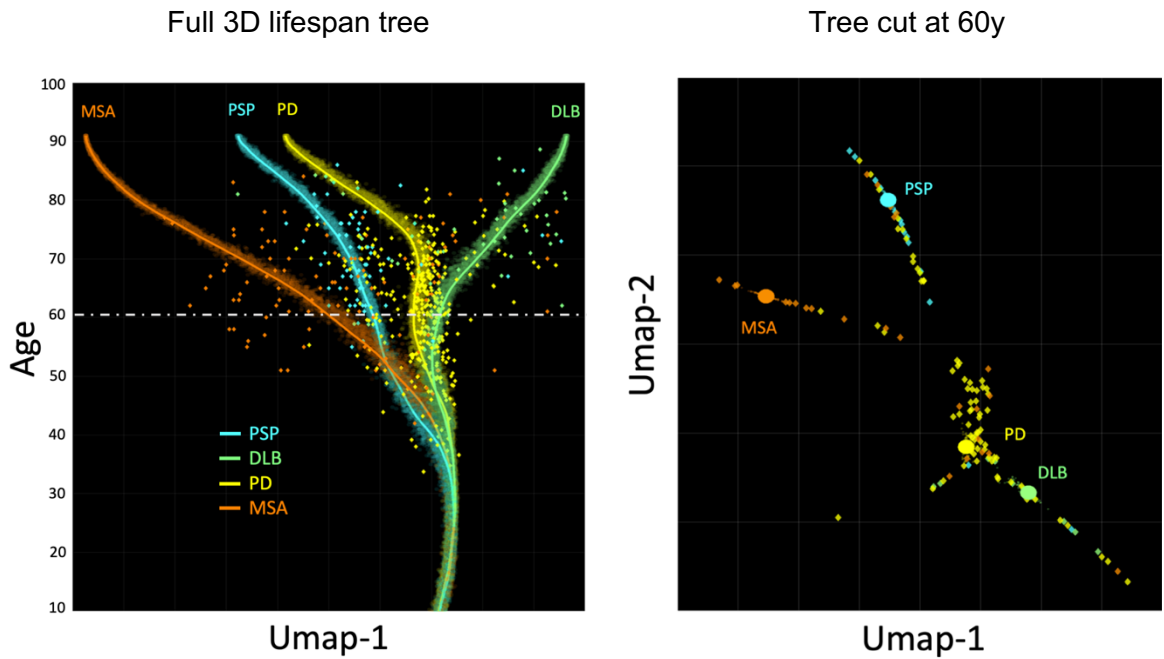

*Figure 4: Left: The 3D motor tree composed by four branches for Parkinson's Disease (PD), progressive supranuclear palsy (PSP), dementia with Lewy bodies (DLB) and multiple system atrophy (MSA). Right: A 2D tree cut at 60y (including test subjects between 58y and 62y). The cloud dots surrounding the branches are the synthetic samples resulting from the Monte Carlo simulation. The diamonds represent the test subjects.*

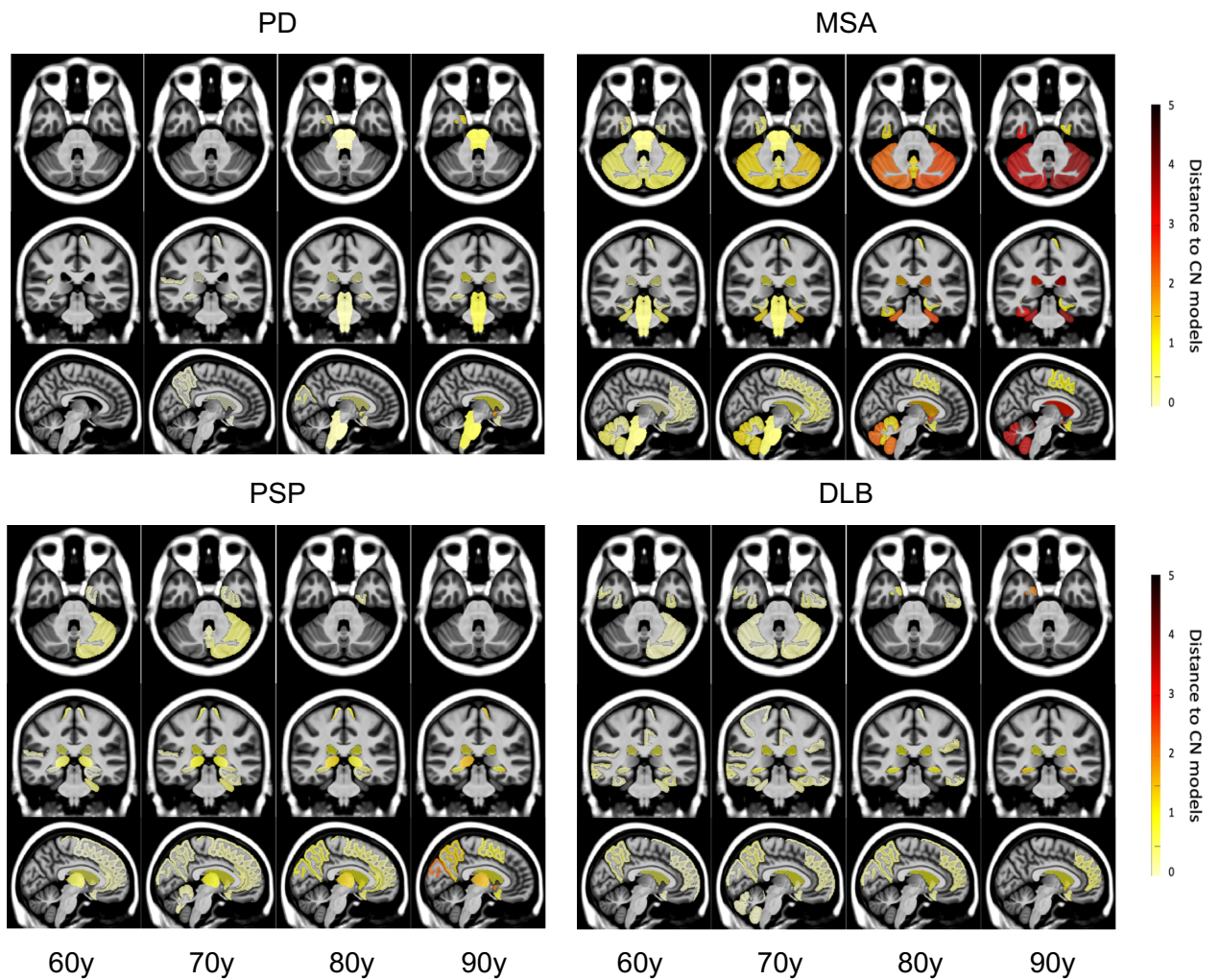

Figure 5: Atlas of the progression of the distance to normal aging (i.e., atrophy severity) for each neurodegenerative diseases reported in the motor tree. The four columns represent the anatomical progression of each disease at 60y, 70y, 80y and 90y. DLB=dementia with Lewy bodies, MSA=multiple system atrophy, PD=Parkinson's Disease, PSP=progressive supranuclear palsy.

*Supplementary Material*
# Lifespan tree of brain anatomy: diagnostic values for motor and cognitive neurodegenerative diseases

Pierrick Coupé[1] (https://orcid.org/0000-0003-2709-3350), Boris Mansencal[1], José V. Manjón[2], Patrice Péran[3], Wassilios G. Meissner[4,5,6] (https://orcid.org/0000-0003-2172-7527), Thomas Tourdias[7,8] (https://orcid.org/0000-0002-7151-6325), Vincent Planche[4]

[1] CNRS, Univ. Bordeaux, Bordeaux INP, LABRI, UMR5800, F-33400 Talence, France
[2] ITACA, Universitat Politècnica de València, 46022 Valencia, Spain
[3] ToNIC, Toulouse NeuroImaging Center, Université de Toulouse, Inserm, UPS, Toulouse, France
[4] CHU Bordeaux, Service de Neurologie des Maladies Neurodégénératives, IMNc, F-33000 Bordeaux, France
[5] Univ. de Bordeaux, CNRS, IMN, UMR 5293, F-33000 Bordeaux, France
[6] Dept. Medicine, University of Otago, Christchurch, New Zealand, and New Zealand Brain Research Institute, Christchurch, New Zealand
[7] Inserm U1215 - Neurocentre Magendie, Bordeaux F-33000, France
[8] Service de neuroimagerie, CHU de Bordeaux, F-33000 Bordeaux.

# 5 Databases description

*Supplementary Table 1: Description of the databases used for the construction (training) of the cognitive and motor trees. The table provides the number of subjects (N) per class, the sex distribution between (female/male) and the mean age [age range]. During our experiments, we used brain MRI from cognitively normal subjects (CN), patients with Alzheimer's disease (AD), dementia with Lewy bodies (DLB), progressive supranuclear palsy (PSP), behavioral variant frontotemporal dementia (bvFTD), progressive non-fluent aphasia (PNFA) and semantic dementia (SD), Parkinson's disease (PD) and multiple system atrophy (MSA). The links to dataset websites are provided in the acknowledgment section of the paper and the references when available in the method section of the paper.*

| Training DATASETS N= 37594 | AD N = 461 | CN N = 36527 | DLB N = 126 | bvFTD N = 155 | PNFA N = 41 | SD N = 39 | MSA N = 21 | PD N = 133 | PSP N = 91 |
|---|---|---|---|---|---|---|---|---|---|
| ABIDE | | **492** (84/408) 18 [6 - 52] | | | | | | | |
| ADHD200 | | **544** (263/281) 12 [7 - 26] | | | | | | | |
| ADNI | **332** (151/181) 75 [55 – 91] | **404** (203/201) 75 [60 - 90] | | | | | | | |
| ALLFTD | **5** (4/1) 67 [60 - 71] | | **1** (0/1) 66 [66 - 66] | **155** (58/97) 64 [40 - 83] | **41** (22/19) 69 [ 55 - 82] | **39** (16/23) 64 [ 50 - 79] | | **2** (0/2) 62 [55 - 69] | **68** (35/33) 68 [52 - 80] |
| AOMIC | | **1361** (731/630) 22 [18 - 26] | | | | | | | |
| Calgary | | **267** (117/150) 4 [3 - 7] | | | | | | | |

| | | | | | | | | |
|---|---|---|---|---|---|---|---|---|
| CamCAN | | **653** (330/323) 55 [18 - 89] | | | | | | |
| C-MIND | | **236** (129/107) 8 [1 - 19] | | | | | | |
| DLBS | | **315** (198/117) 55 [ 21 - 89] | | | | | | |
| ICBM | | **294** (142/152) 34 [18 - 80] | | | | | | |
| ISYB | | **213** (155/58) 22 [18 - 30] | | | | | | |
| IXI | | **549** (307/242) 49 [20 - 86] | | | | | | |
| MIRIAD | **46** (27/19) 69 [ 59 - 86] | **23** (11/12) 70 [58 - 86] | | | | | | |
| NDAR | | **382** (174/208) 12 [1 - 50] | | | | | | |
| OASIS | **45** (29/16) 77 [63 - 96] | **298** (187/111) 45 [18 - 94] | | | | | | |
| PDBP | | **22** (13/9) 70 [65 - 85] | **125** (22/103) 69 [50 - 90] | | | **21** (4/17) 65 [53 - 80] | **37** (14/23) 62 [40 - 77] | **21** (12/9) 68 [60 - 80] |
| PIXAR | | **155** (84/71) 11 [3 - 39] | | | | | | |
| SALD | | **494** (307/185) 45 [19 - 80] | | | | | | |
| **SLIM** | | **574** (320/254) <br> 20 [17 - 27] | | | | | | |
| **UKB** | **33** (18/15) <br> 70 [52 - 77] | **29251** (14334/14917) <br> 64 [44 - 82] | | | | | **94** (34/60) <br> 70 [55 - 81] | **2** (1/1) <br> 73 [71 - 75] |

*Supplementary Table 2: Description of the databases used for the testing of the cognitive and motor trees. Testing come from totally independent databases from training. The table provides the number of subjects (N) per class, the sex distribution between (female/male) and the mean age [age range]. During our experiments, we used brain MRI from cognitively normal subjects (CN), patients with Alzheimer's disease (AD), dementia with Lewy bodies (DLB), progressive supranuclear palsy (PSP), behavioral variant frontotemporal dementia (bvFTD), progressive non-fluent aphasia (PNFA) and semantic dementia (SD), Parkinson's disease (PD) and multiple system atrophy (MSA). The links to dataset websites are provided in the acknowledgment section of the paper and the references when available in the method section of the paper*

| Testing DATASETS N= 1754 | AD N = 488 | CN N = 528 | DLB N = 47 | bvFTD N = 90 | PNFA N = 40 | SD N = 44 | MSA N = 117 | PD N = 333 | PSP N = 67 |
|---|---|---|---|---|---|---|---|---|---|
| 4RTNI | | | | | | | | | **67** (38/29) 71 [55 - 86] |
| AIBL | **47** (29/18) 73 [55 - 93] | **232** (120/112) 72 [60 - 89] | | | | | | | |
| FMSA | | | | | | | **117** (60/57) 66 [42 - 84] | | |
| NACC | **441** (275/166) 74 [46 - 96] | **161** (112/49) 73 [30 - 100] | **47** (8/39) 74 [55 - 89] | **20** (9/11) 65 [59 - 76] | **6** (4/2) 68 [57 - 77] | **7** (5/2) 67 [55 - 85] | | | |
| NIFD | | **135** (76/59) 63 [39 - 81] | | **70** (22/48) 62 [45 - 74] | **34** (19/15) 69 [54 - 81] | **37** (15/22) 63 [50 - 73] | | | |
| PPMI | | | | | | | | **333** (120/213) 62 [34 - 83] | |

# 6 Confusion Matrices

In this supplementary section, we provide the confusion matrices related to the cognitive and motor trees experiments.

| | Lifespan tree | | | | | | | | SVM on volumes | | | | | | |
|---|---|---|---|---|---|---|---|---|---|---|---|---|---|---|---|
| | AD | CN | DLB | PNFA | PSP | SD | bvFTD | | AD | CN | DLB | PNFA | PSP | SD | bvFTD |
| AD | 214 | 106 | 65 | 2 | 12 | 35 | 54 | AD | 212 | 248 | 5 | 4 | 5 | 1 | 13 |
| CN | 21 | 390 | 57 | 18 | 28 | 1 | 13 | CN | 9 | 514 | | 1 | 3 | | 1 |
| DLB | 13 | 9 | 15 | 1 | 3 | 2 | 4 | DLB | 15 | 29 | | | 1 | | 2 |
| PNFA | 2 | 4 | 6 | 13 | 6 | 3 | 6 | PNFA | 1 | 20 | | 7 | 5 | 3 | 4 |
| PSP | 2 | 5 | 4 | 5 | 48 | | 3 | PSP | 1 | 23 | | 4 | 39 | | |
| SD | | 1 | 1 | | | 40 | 2 | SD | 7 | 6 | | | | 25 | 6 |
| bvFTD | 3 | 10 | 9 | 8 | 7 | 14 | 39 | bvFTD | 7 | 29 | | 2 | 3 | 3 | 46 |

*Supplementary Figure 1: Confusion matrices obtained for the cognitive neurodegenerative diseases experiment.*

| | Lifespan tree | | | | | SVM on volumes | | | |
|---|---|---|---|---|---|---|---|---|---|
| | DLB | MSA | PD | PSP | | DLB | MSA | PD | PSP |
| DLB | 31 | 4 | 9 | 3 | DLB | 21 | 3 | 13 | 10 |
| MSA | 18 | 49 | 31 | 19 | MSA | 17 | 32 | 49 | 19 |
| PD | 80 | 21 | 189 | 43 | PD | 64 | 7 | 225 | 37 |
| PSP | 5 | 2 | 6 | 54 | PSP | 3 | 1 | 5 | 58 |

*Supplementary Figure 2: Confusion matrices obtained for the motor diseases experiment.*

# 7 Diagnosis and prognosis of Alzheimer's disease

In this supplementary experiment, we validated the proposed method on the binary classification problems of AD diagnosis and prognosis to compare them with existing state-of-the-art methods. For this experiment, we used the same databases as in (Wen et al. 2020; Coupé et al. 2022; H.-D. Nguyen, Clément, Mansencal, et al. 2023).

## 7.1 Comparation with state-of-the-art methods

During this experiment, we compared our method with the hippocampal-amygdalo-ventricular atrophy score (HAVAs) proposed in (Coupé et al. 2022) (HAVAs is freely available at [www.volbrain.net](www.volbrain.net)). HAVAs is based on a similar strategy to lifespan tree of the brain anatomy but relies on a manual selection of key structures (i.e., hippocampus, amygdala, and inferior ventricles). This manual selection needs to be adapted for each scenario, and it might be impractical for multi-class classification where pathologies partially share anatomical atrophic patterns. The comparison of lifespan tree of the brain anatomy and HAVAs was performed to verify that the UMAP correctly learned discriminative features during training. By validating the performance of lifespan tree of the brain anatomy against HAVAs, we aimed to highlight the robustness and flexibility of our proposed method in various differential diagnosis tasks. This comparison underscores the effectiveness of our approach in automatically and accurately distinguishing between different neurological conditions.

Second, as shown in (Wen et al. 2020), many proposed deep learning methods suffer from data leakage, resulting in biased reported performances. Additionally, most published studies use the same dataset for both training and testing, which leads to over-optimistic performance evaluations (Wen et al. 2020; Bron et al. 2021; Varoquaux 2018). Consequently, we decided to report the scores of the well-evaluated methods proposed in (Wen et al. 2020) as state-of-the-art deep learning methods, as the training was well-designed and the proposed methods were well-validated on external datasets. We selected three end-to-end deep learning methods from (Wen et al. 2020):

- ROI-based Convolutional Neural Network (CNN): This method focuses on the hippocampal area, which is crucial for AD diagnosis.
- Subject-based CNN: This method uses the entire brain image for processing.
- Patch-based CNN: This method processes the whole image patch by patch, ensuring a detailed analysis of different brain regions.

Moreover, (Wen et al. 2020) demonstrated that classical machine learning methods (e.g., SVM) often perform better than deep learning methods for AD diagnosis. Therefore, we included their SVM on gray matter (GM) density method in our comparison. These four strategies represent the state-of-the-art frameworks for AD detection and prognosis effectively.

Finally, we also included a more recent hybrid method that combines deep learning and machine learning, as described by (H.-D. Nguyen, Clément, Mansencal, et al. 2023). This method involves estimating deep grading features from MRI intensity, which are then fed into a classifier—in this case, a graph convolutional network (GCN). However, to focus on the efficiency of the features themselves rather than the classifier, we used the results based on SVM instead of GCN, as originally proposed. This hybrid approach leverages the strength of deep learning in feature extraction and the robustness of classical machine learning classifiers, providing a comprehensive comparison against our proposed Lifespan tree method. By including this method, we aimed to highlight the effectiveness of different strategies in handling MRI-based differential diagnosis and to showcase the versatility and robustness of our Lifespan tree approach in comparison to state-of-the-art methodologies.

In summary, our comparisons encompass a diverse array of techniques, including deep learning methods, classical machine learning, and hybrid approaches. This comprehensive evaluation ensures a thorough assessment of our lifespan tree of the brain anatomy method's performance in the context of current advanced diagnostic frameworks.

## 7.2 Results

We first validated the proposed method on the binary classification problems of AD diagnosis and prognosis to compare them with existing state-of-the-art methods. For this experiment, we used the same databases as in (Wen et al. 2020; Coupé et al. 2022; H.-D. Nguyen, Clément, Mansencal, et al. 2023). Our training set comprised 377 AD and 702 cognitively normal (CN) individuals older than 55 years from the ADNI and OASIS databases. For testing, we included 47 AD and 232 CN individuals from the AIBL dataset and 490 individuals with mild cognitive impairment (MCI) from the ADNI dataset, following the setup in (Wen et al. 2020; Coupé et al. 2022; H.-D. Nguyen, Clément, Mansencal, et al. 2023). The AD and CN subjects were from AIBL, while the MCI subjects were from ADNI. The MCI group was further split into stable MCI (sMCI), who remained stable over three years, and progressive MCI (pMCI), who converted to AD within 36 months following the baseline visit.

*Supplementary Table 3: Comparison with state-of-the-art for AD diagnosis and prognosis on external databases. Balanced accuracy (Balanced ACC) in % is provided for each method for both datasets. For Deep Grading, CNN-based methods and HAVAs, the results published on the same datasets (Wen et al. 2020; Coupé et al. 2022; H.-D. Nguyen, Clément, Mansencal, et al. 2023) are used.*

| Balanced ACC in % | AIBL (232 CN / 47 AD) | ADNI (255 sMCI / 235 pMCI) |
|---|---|---|
| **Lifespan tree** | 88 | 73 |
| HAVAs (Coupé et al. 2022) | 88 | 73 |
| SVM on deep grading (H.-D. Nguyen et. al. 2023) | 89 | 69 |
| **SVM on volumes** | 84 | 72 |
| SVM on GM density (Wen et al. 2020) | 88 | 67 |
| ROI-based CNN (Wen et al. 2020) | 84 | 70 |
| Subject-based CNN (Wen et al. 2020) | 83 | 69 |
| Patch-based CNN (Wen et al. 2020) | 81 | 70 |

Supplementary Table 3 presents the results for AD diagnosis (AD vs. CN) and AD prognosis (pMCI vs. sMCI) tasks on external datasets, measured in terms of balanced accuracy (BACC). In both tasks, all methods were trained exclusively on AD and CN subjects, with the AD prognostic task demonstrating the methods' generalization capabilities on unseen tasks.

For AD diagnosis, the SVM on deep grading method (H.-D. Nguyen, Clément, Mansencal, et al. 2023) achieved the highest BACC at 89%, followed by the lifespan tree of the brain anatomy, HAVAs (Coupé et al. 2022) and SVM on GM density (Wen et al. 2020), each with 88% BACC (see supplementary Table 4 for detailed results). The end-to-end deep learning methods including ROI-based CNN, Subject-based CNN, and Patch-based CNN, produced the lowest performances, ranging from 81% to 84%.

*Supplementary Table 4: Comparison of the proposed methods on external datasets for AD diagnosis on external databases. During training, we used 377 AD and 702 CN older than 55y for training from ADNI and OASIS databases. In addition, we used 1874 CN younger than 55y for building lifespan trajectories as detailed in (Coupé et al. 2022). During testing, we used 47 AD and 232 CN from AIBL.*

| Classification on external datasets in % | BACC | SEN. AD | SEN. CN |
|---|---|---|---|
| **Lifespan tree** | 88 [84-92] | 84 [75-92] | 92 [90-94] |
| **SVM on volumes** | 84 [80-89] | 79 [70-87] | 98 [97-99] |

Furthermore, lifespan tree of the brain anatomy and HAVAs yielded identical results, reinforcing that UMAP effectively extracts discriminative features. While utilizing all structures (i.e., lifespan tree of the brain anatomy) slightly reduced the results, this approach remains a viable alternative.

For AD prognosis at 3 years, the lifespan tree of the brain anatomy and HAVAs methods achieved the highest BACC at 73% (see supplementary Tab. 5 for detailed results). Similar to the AD diagnosis task, the lifespan tree of the brain anatomy and HAVAs methods produced identical results. In this task, the proposed SVM-based methods also performed competitively, with a BACC of 72%. The lowest accuracy was observed with the SVM on GM density method (Wen et al. 2020), despite its strong performance in the AD diagnosis task.

Overall, this comparison demonstrates that the proposed lifespan tree of the brain anatomy delivers state-of-the-art results for these binary classification tasks. More importantly, the lifespan tree of the brain anatomy offers a significant advantage in providing easily and visually interpretable results. As shown in supplementary Figure 3, a new subject can be directly classified using the closest branch. The distance to branches provides an indication of how well a test subject fits within a class. Finally, the dispersion of synthetic sample clouds surrounding the branches illustrates the overlap between class domains with respect to age.

*Supplementary Table 5: Comparison of the proposed methods on external datasets for AD prognosis on external databases. During training, we used 377 AD and 702 CN older than 55 for training coming from ADNI and OASIS databases. During testing, we used 235 pMCI and 255 sMCI from ADNI. The best results are indicated in red and second best in green.*

| Classification on external datasets in % | BACC | SEN. pMCI | SEN. sMCI |
|---|---|---|---|
| Lifespan tree | 73 [69-77] | 72 [66-77] | 74 [68-79] |
| SVM on volumes | 72 [68-75] | 65 [58-71] | 80 [74-84] |

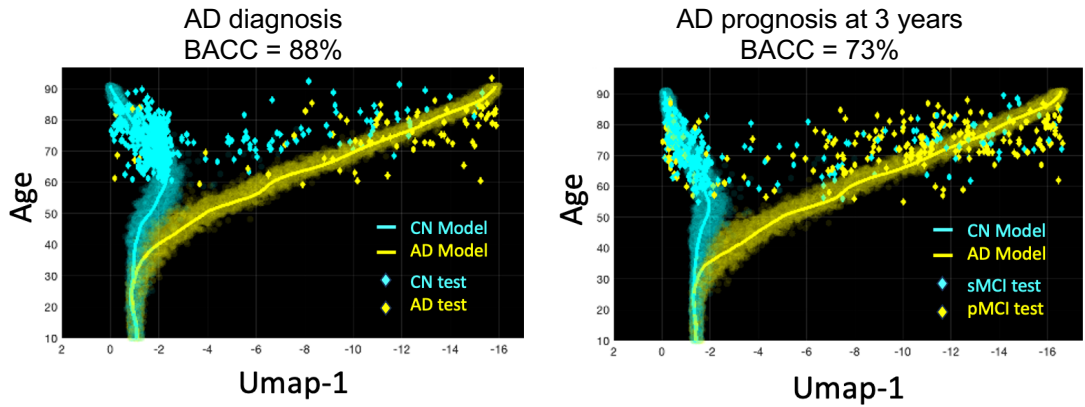

*Supplementary Figure 3: 3D Lifespan tree composed by two branches, one for AD and one for CN. The cyan branch represents the CN model surrounded by the cloud of synthetic samples resulting from the Monte Carlo simulation. The yellow branch represents the AD model surrounded by the cloud of synthetic samples resulting from the Monte Carlo simulation. Diamonds represent the testing subjects projected into the tree space. Left: the testing subjects are the AD and CN from AIBL. Right: the testing subjects are the sMCI and pMCI from ADNI.*

# 8  Binary classifications

In this section we provide the results of binary classifications between populations involved into cognitive and motor trees experiments.

*Supplementary Table 6: Results of binary classification task on cognitive disorders using lifespan tree method. For training, we used 31181 CN older than 44y, 461 AD, 155 bvFTD, 39 SD, 41 PNFA, 91 PSP and 126 DLB (see Table 1). For testing on external datasets, we used 528 CN, 488 AD, 90 bvFTD, 44 SD, 40 PNFA, 67 PSP-RS and 47 DLB (see Table 1).*

| Classification on external datasets | BACC | SEN. | SPE. |
|---|---|---|---|
| **AD vs. CN** | 80 [77-82] | 76 [72-79] | 84 [81-87] |
| **bvFTD vs. CN** | 79 [74-84] | 70 [60-79] | 89 [86-91] |
| **SD vs. CN** | 96 [92-98] | 96 [89-100] | 96 [94-97] |
| **PNFA vs. CN** | 82 [75-87] | 83 [70-93] | 80 [77-84] |
| **PSP vs. CN** | 79 [73-84] | 75 [64-85] | 83 [80-86] |
| **DBL vs. CN** | 74 [67-80] | 70 [57-83] | 77 [73-80] |

*Supplementary Table 7: Results of binary classification task on motor disorders using our lifespan tree method. For training, we used 91 PSP, 126 DLB, 133 PD and 21 MSA (see Table 1). For testing on external datasets, we used 67 PSP, 47 DLB, 333 PD and 117 MSA (see Table 1).*

| Classification on external datasets | BACC | SEN. | SPE. |
|---|---|---|---|
| **PSP vs. PD** | 84 [79-88] | 90 [82-96] | 78 [74-82] |
| **DLB vs. PD** | 75 [69-81] | 81 [69-92] | 69 [64-74] |
| **MSA vs. PD** | 72 [66-76] | 62 [53-71] | 81 [77-85] |